\documentclass[aps,prb,longbilbiography,twocolumn,nofootinbib]{revtex4-1}
\usepackage{amsfonts}
\usepackage{amsmath,latexsym}
\usepackage{amssymb,amsmath,amsthm}
\usepackage[dvips]{graphicx}
\usepackage{verbatim} %
\usepackage{color}
\usepackage[normalem]{ulem}
\usepackage{textgreek}
\newcommand{\beq}{\begin{equation}}
\newcommand{\eeq}{\end{equation}}
\newcommand{\bea}{\begin{eqnarray}}
\newcommand{\eea}{\end{eqnarray}}
\usepackage{bm}
\usepackage{color}

\newcommand{\bma}{\left(\begin{matrix}}
\newcommand{\ema}{\end{matrix}\right)}

\newcommand{\om}{\omega}
\def\build#1_#2{\mathrel{\mathop{#1}\limits_{#2}}}
\definecolor{pink}{rgb}{1,0.5,0.5}
\definecolor{violet}{rgb}{1,0,1} 
\definecolor{red}{rgb}{1,0,0}
\definecolor{yellow}{rgb}{0.7,1,0}
\definecolor{orange}{rgb}{1,0.5,0}
\definecolor{white}{rgb}{1,1,1}
\definecolor{blue}{rgb}{0,0,1}
\definecolor{cyan}{rgb}{0,1,1}
\begin{document}

\begin{abstract}
{In this work we present an implementation of the analytical string theory recently applied to the description of glasses. These are modeled as continuum media with embedded elastic string heterogeneities, randomly located and randomly oriented, which oscillate around a straight equilibrium position with a fundamental frequency depending on their length. The existence of a length distribution reflects then in a distribution of oscillation frequencies which is responsible for the Boson Peak in the glass density of states. Previously, it has been shown that such a description can account for the elastic anomalies reported at frequencies comparable with the Boson Peak: the strong phonon scattering and the negative dispersion in the sound velocity, as a result of the interference of the string oscillations with propagating sound plane waves.  
Here we start from the generalized hydrodynamics to determine the dynamic correlation function $S(k,\om)$ associated with the coherent, dispersive and attenuated, sound waves resulting from such interference. 
We show that, once the vibrational density of states has been measured, we can use it for univocally fixing the string length distribution inherent to a given glass. The density-density correlation function obtained using such distribution is strongly constrained, and able to account for the experimental data collected on two prototypical glasses with very different microscopic structure and fragility: glycerol and silica.
The obtained string length distribution is compatible with the typical size of elastic heterogeneities previously reported for silica and supercooled liquids, and the atomic motion associated to the string dynamics is consistent with the soft modes recently identified in large scale numerical simulations as non-phonon modes responsible for the Boson Peak. The theory is thus in agreement with the most recent advances in the understanding of the glass specific dynamics and offers an appealing simple understanding of the microscopic origin of the latter, while raising new questions on the universality or material-specificity of the string distribution properties.
}

\end{abstract}

\title{Elastic anomalies in glasses: the string theory understanding in the case of Glycerol and Silica}

\author{Ernesto Bianchi$^{1,2}$, Valentina M. Giordano$^3$ and Fernando Lund$^{1,4}$}
\affiliation{$^1$Departamento de F\'\i sica, Facultad de Ciencias F\'\i sicas y Matem\'aticas, Universidad de Chile, Santiago, Chile \\
$^2$Instituto de F\'\i sica, Pontificia Universidad Cat\'olica de Chile, Casilla 306, Santiago, Chile \\
$^3$Institut Lumi\`ere  Mati\`ere, UMR 5306 Universit\'e Lyon 1-CNRS,  F-69622 Villeurbanne Cedex, France \\
$^4$CIMAT, Facultad de Ciencias F\'\i sicas y Matem\'aticas, Universidad de Chile, Santiago, Chile}

\date{ \today}

\maketitle

\section{Introduction}

A long standing issue in condensed matter physics is represented by the understanding of the low frequency vibrational properties of glasses. It is well known that at long wavelengths and low frequencies, the continuum description of elastic properties in a solid generally works well in a crystal, which means that the Debye approximation of the density of states is a good description of the actual situation~\cite{kittel}.
This approximation however dramatically fails when it comes to a disordered material such as a glass, in spite of intuition that says this should not happen. Indeed, on the macroscopic scale, the wavelengths of sound waves are much larger than the typical lengthscale of the disorder, thus the microscopic structure should be irrelevant, and a description of the medium as a continuum should work well. 
We can expect this approximation to fail when the wavelength becomes comparable with the microscopic structure. It is however found to fail much earlier than expected, at wavelengths on the order of tens of mean interatomic distances, a mesoscopic scale corresponding to the medium range order in glasses and at which it still works reasonably well in crystals\cite{Monaco2009}. 

Historically, the above failure has been observed through the deviation of the vibrational density of states from the Debye prediction at frequencies of few THz, about 1/10 of the Debye frequency, pointing to the existence of an excess of soft modes, called the Boson Peak~{\cite{Buchenau1984}}. Such anomaly in the vibrational density of states can be connected with the anomalous temperature dependence of the specific heat at low temperature, as well as to the existence of a plateau in the thermal conductivity, with the suppression of the Umklapp peak typical of crystals, at temperatures of about 10~K~\cite{Zeller1971,Courtens2001}. Such excess is qualitatively similar for many such materials, and the details, but not the broad features, depend on external parameters such as temperature, density, pressure, as well as chemical and thermal history \cite{Wischnewski1998,Caponi2007,Inamura2000,Zanatta2010,Monaco2006,Orsingher2010,Niss2007,Hong2008,Caponi2009,Monaco2006b}.

This anomaly in the density of states is accompanied by some so-called elastic anomalies in the longitudinal acoustic phonons, as evidenced by a series of experimental studies~\cite{Monaco2009,Baldi2010,Baldi2011,Ruta2012} : a negative dispersion of the sound velocity, with a minimum close to the
Boson Peak frequency, and a strong attenuation $\Gamma$ of the acoustic phonons, whose dependence on the wavevector $k$ has long been  believed to be compatible with a Rayleigh scattering mechanism ($\Gamma \propto k^{d+1}$, where $d$ is the dimensionality of space), but recently found to go like $-k^{d+1} ln k$ near the Boson Peak frequency~\cite{Gelin2016,Caroli2019}. 
As such, the failure of the Debye approximation seems to be drawn by a strong scattering of sound waves, whose origin is still a matter of debate and which leads to a cross-over from well-defined acoustic modes for long wavelengths to ill-defined ones for smaller wavelengths, the so-called Ioffe-Regel crossover. Three dynamical regimes have thus been identified in glasses~\cite{Beltukov2016,Tanguy2002}: i) a low-frequency plane-wave dominated regime, where phonons undergo weak scattering and are equivalent to phonons in crystals ii) a strong scattering regime, above the Ioffe-Regel limit, where the corresponding phonons are called ``diffusons'' and cannot be thought of as propagative plane waves anymore~\cite{AllenFeldman1999}, and iii) the Anderson localization regime near the mobility edge~\cite{AllenFeldman1993,Ludlam2003,Beltukov2016}.  The existence of these three regimes strongly impacts transport properties, so that their microscopic understanding has become essential for many technological applications aiming at using glasses for thermal insulation or confinement.

{Understanding the microscopic origin of the strong scattering regime, and thus of elastic anomalies and Boson Peak, represents a challenge that is at the focus of an intense theoretical and experimental research effort since at least 50 years. Many theories and models have been developed\cite{Parshin1994,Gotze2000,Schirmacher1998,Schirmacher2006,Schirmacher2007,Grigera2003,Ganter2010,Beltukov2013}, which can be divided in three main groups: i) the existence of quasi-local vibrational states, produced by soft anharmonic potentials~\cite{Buchenau1991,Gurevich1998,Gurevich2003}, ii) the existence of nanometric elastic heterogeneities~\cite{Schirmacher2006,Duval2007,Schirmacher1998,Gotze2000,Kantelhardt2001,Schirmacher2007}, and iii) the identification of the Boson Peak with the first transverse Van Hove singularity, broadened by disorder and shifted because of the density difference with respect to the corresponding crystalline phase~\cite{Taraskin2001,Chumakov2011}.

Recent work has shown that the theory of elastic nano-heterogeneities is unable to reproduce the observed over-Rayleigh acoustic attenuation~\cite{Caroli2019}. On the other hand, recent large-scale numerical simulations have unveiled the presence of soft quasi-localized non-phononic modes, which would coexist with normal acoustic phonons~\cite{Shimada2018,Lerner2016,Gartner2016,Kapeteijns2018,Lerner2018} and characterized by a particle displacement decreasing like $r^{-2}$ in 3D, where $r$ is the distance away from the center of the mode. The low-frequency density of states is thus the result of the sum of a Debye contribution for the acoustic phonons and a non-phononic contribution coming from these modes and going as $\omega^4$.
These soft modes would stem out from microscopic structural arrangements and/or heterogeneities in the glass and can be associated to another class of theoretical models, which has focused on the identification of string-like behavior in glasses.  Cooperative string-like motion was successively found in super-cooled liquids~\cite{Donati1998,Zhang2015} and Yu et al. \cite{Yu2013} proposed an interpretation of the $\beta$-relaxation in a variety of supercooled liquids and glasses in terms of string-like configurations.
Concerning glasses, already in the 90's Schober et al.~\cite{Schober1993} in a numerical simulation of a soft-sphere glass identified well below the glass transition temperature vibrational modes involving only atoms arranged in a string-like pattern. 
Novikov and Surotsev \cite{Novikov1999} showed that Raman scattering data from glasses could indeed be explained by vibrational eigenmodes localized along a one-dimensional spatial geometry.  More recently, Concustell et al. \cite{Concustell2011} have induced and characterized elastic anisotropy in a bulk metallic glass, showing that their findings are consistent with an alignment of string-like atomic arrangements. }

{Recently, one of us has developed an analytical description showing that the presence of strings in a glass thought as a continuum medium can give rise to the vibrational anomalies observed in glasses~\cite{Lund2015}. Here we develop this theory and provide a formula to directly link the density of states $g(\om)$ of an amorphous material in the THz range with its longitudinal dynamic structure factor $S(k,\om)$. We find that the string distribution obtained from the experimental Boson Peak is able to nicely fit the inelastic x ray scattering data on the $S(q,\om)$ obtained on glycerol ~\cite{Monaco2009} and silica~\cite{Baldi2010} with an agreement as good as the one historically obtained fitting the same data with a Damped Harmonic Oscillator (DHO) model.  }

This paper is organized as follows: Section II shows how to compute the dynamic structure factor $S(\vec k, \om)$ for coherent acoustic waves propagating in a medium described by a complex index of refraction, starting from the Boson peak. This computation is done following our previous work\cite{Lund2015} and relies on the properties of an acoustic wave propagating in a medium filled with elastic strings randomly placed and oriented \cite{Lund2015,Churochkin2016}. For the reader's convenience, some details of the index of refraction computation are reproduced in Appendix \ref{Apa}. The relation of the current approach with the Damped Harmonic Oscillator (DHO) model is discussed in Appendix \ref{sec:DHO}. {Section III is devoted to the application of the model for fitting experimental data on two prototypical glasses,  glycerol and silica. The discussion of the results of such fitting, and notably the obtained string length distribution and its relation to the elastic anomalies is reported in Section IV A, while Section IV B focuses on the comparison of our results and recent numerical simulations, specifically concerning the atomic motion in the strings implicit to the our model. Finally, Section V presents our concluding remarks.}

\section{The string model}
\label{Sec:string_model}
Here we recall the basics of the model developed in \cite{Lund2015,Churochkin2016}. It is based on continuum mechanics. The glass is modeled as a Debye solid plus an embedded distribution of elastic strings. These strings are pinned at their ends and have a distribution of lengths. They can oscillate around an equilibrium position and their fundamental mode of vibration gives rise to the excess modes over the Debye level observed in many glasses. Also, they interact with elastic waves, {leading to the frequency-dependence of phase velocity and attenuation of coherent sound waves}

\subsection{Preliminaries}
We consider then an elastic, homogeneous, isotropic solid with Lam\'e coefficients $\lambda,\mu$, mass density $\rho$ and number density $n$. The displacement at time $t$ of a point whose equilibrium position is $\vec x$ is denoted $\vec u(\vec x,t)$, and its velocity is $\vec v(\vec x,t) \equiv \partial \vec u(\vec x,t) /\partial t $. The current $\vec j \equiv n\vec v$ has a longitudinal component $j(\vec x,t)$ that, in the absence of strings, obeys the equation 
\begin{equation}
\label{eq:we_nostrings}
 -\frac{1}{c_L^2}\frac{\partial^2 j(\vec{x},t)}{\partial t^2} + \nabla^2 j(\vec{x},t) + \xi\nabla^2\frac{\partial j(\vec{x},t)}{\partial t} = 0
\end{equation}
where $c_L^2 = \rho/(\lambda +2\mu)$ and $\xi$ is a phenomenological damping coefficient \cite{Landau1970}. Looking for traveling waves along one dimension
\beq
j(\vec x,t) = j_0 e^{i(\om t -kx)}
\eeq
leads to the dispersion relation
 \beq
 \label{eq:baredisprel}
 \frac{\om^2}{c_L^2} - k^2 -ik^2\om \xi = 0 \, .
 \eeq
and, when $\om \xi \ll 1$, we have 
\beq
j(\vec x,t) = j_0 e^{i(\om t- \om x/c_L)} e^{-x (\om^2 \xi)/2c_L}  \, ,
\eeq
the well-known result that, at low frequencies, elastic waves have a damping proportional to frequency squared, or that there is an attenuation length proportional to $\om^{2}$. For later purposes we introduce the (``bare'') Green's function ${g}_0 (\vec r,\vec r_0, t-t_0)$, the solution of 
\bea
\label{nos2}
 -\frac{1}{c_L^2} {\ddot g}_{0} (\vec r,\vec r_0, t-t_0) +\nabla^2 {g}_{ 0 }(\vec r,\vec r_0, t-t_0)  \hspace{5em} \\ 
 + \xi \, \nabla^2 {\dot g}_{ 0}  (\vec r,\vec r_0, t-t_0) = - \delta (\vec r-\vec r_0) \delta (t-t_0)  \nonumber
\eea
where an overdot denotes differentiation with respect to time. We shall use the definition of the Fourier transform $\tilde F(\vec k,\om)$ of a function $F(\vec x,t)$ as
\beq
\tilde F(\vec k,\om) \equiv \int_{-\infty}^{\infty} d^3x \, dt\,  e^{i(\vec k \cdot \vec x - \om t)} F(\vec x,t)
\eeq
so the bare Green's function in frequency-wavenumber space is given by
\begin{equation}
\label{eq:baregreen}
    \tilde g_0 (\vec{k},\omega) = \frac{-1}{(\omega^2/c_L^2)-k^2-ik^2\xi\omega}
\end{equation}
whose poles reproduce the dispersion relation (\ref{eq:baredisprel}).

If now we consider the solid to be in thermodynamic equilibrium at temperature $T$, the currents will be fluctuating quantities. Of particular interest is the dynamic correlation
\beq
\label{eq:defJ}
J(\vec{k},t) \equiv \langle j^*(\vec{k},0)j(\vec{k},t) \rangle
\eeq
where the brackets denote a thermal average. Also, we shall drop the tildes from the Fourier transforms to simplify the notation. Note that, because of thermal equilibrium this is a real function that is defined for positive times $t$, and it can be extended to negative times imposing that it be even in time. Of particular interest is the Fourier transform
\bea
\label{eq:Jkw}
J(\vec k, \om )  &\equiv & 2\text{Re} \left[ {\cal J} \right] . \\
 {\cal J}  & \equiv & \int_0^{\infty} dt e^{-i\om t} J(\vec k, t) 
\eea
A standard treatment \cite{MF} of the initial value problem of Eqns. (\ref{eq:we_nostrings}) and (\ref{nos2}) leads to an expression of the bare correlation $J_0$ in terms of the bare Green's function $g_0$:
\bea
\label{eq:barecorrel1}
J_0(\vec k, \om ) & = & -2\frac{J_0(\vec{k},0)}{c_L^2}\omega \text{Im}\left[g_0(\vec{k},\omega)\right] \\
& = &  \frac{2}{c_L^2} J_0(k,0) \frac{\xi \om^2 k^2}{\left(\frac{\om^2}{c_L^2} -k^2\right)^2 + \xi^2 \om^2 k^4}
\eea
where the second line is a well known expression \cite {BY}, that has been obtained here substituting (\ref{eq:baregreen}) into (\ref{eq:barecorrel1}).

\subsection{Addition of strings}
\label{sec:addstrings}
We now consider elastic waves as above, propagating in a medium with a frequency dependent index of refraction induced by the presence of strings:
\bea
\label{eq:we}
    -\frac{1}{c_L^2}\frac{\partial^2 j(\vec{x},t)}{\partial t^2} + \nabla^2 j(\vec{x},t) + \xi\nabla^2\frac{\partial j(\vec{x},t)}{\partial t} && \\
     - \nabla^2\left[f(t)*j(\vec{x},t)\right] &= & 0 \nonumber
\eea
where $*$ means convolution in time and $f(t)$ must vanish for negative times due to causality. The road to Eqn. (\ref{eq:we}) as well as its relation to strings is recalled in Appendix \ref{Apa}. There is a Green's function associated with this equation, the solution to 
\bea
    -\frac{1}{c_L^2}\frac{\partial^2 g(\vec{x},\vec{x_0},t-t_0)}{\partial t^2} + \nabla^2 g(\vec{x},\vec{x_0},t-t_0)&& \nonumber \\ + \xi\nabla^2\frac{\partial g(\vec{x},\vec{x_0},t-t_0)}{\partial t} - \nabla^2\left[f(t)*g(\vec{x}-\vec{x_0},t-t_0)\right] \nonumber \\ = - \delta(\vec{x}-\vec{x_0})\delta(t-t_0) \nonumber \\
\eea
which gives, in frequency-wavenumber space,
\begin{equation}
    g(\vec{k},\omega) = \frac{-1}{(\omega^2/c_L^2)-k^2-ik^2\xi\omega+k^2f(\omega)}
    \label{eq:fullgxi}
\end{equation}
whose poles give an effective, complex, wavenumber as a function of frequency, say $K(\om)$, that defines a phase velocity, $v \equiv \om/\text{Re}[K(\om)] $ and an attenuation length $\ell \equiv -1/\text{Im} [K(\om)]$. Alternatively, both the phase velocity and the attenuation length can be expressed as functions of the wavenumber $k$.

Consider now Eqn. (\ref{eq:we}) in wavenumber space:
\bea
    -\frac{1}{c_L^2}\frac{\partial^2 j(\vec{k},t)}{\partial t^2} - k^2 j(\vec{k},t) - k^2\xi\frac{\partial j(\vec{k},t)}{\partial t} & & \\
     + k^2\int_{-\infty}^{\infty}dt'\ f(t')j(\vec{k},t-t') &=& 0 \nonumber \, .
     \label{eq:wek}
\eea
Multiplying this equation by $j^*(\vec k,0)$, {and} taking the thermal average, {we get
\bea
\label{eq:12EB}
    \frac{-1}{c_L^2}\frac{\partial^2 J(\vec{k},t)}{\partial t^2} - k^2 J(\vec{k},t) - k^2\xi\frac{\partial J(\vec{k},t)}{\partial t} \hspace{2em} &&  \\
    + k^2\int_{-\infty}^{t}dt'\ f(t')J(\vec{k},t-t') +k^2 \langle j^*(\vec k,0) {\cal I} (\vec k, t) \rangle &=& 0 \nonumber
\eea
This equation is similar to, but differs from, the equation obeyed by the dynamic form factor in a memory function formalism\cite{Scopigno2005}. This point is further discussed in Appendix \ref{sec:DHO}.}
Further multiplying by $\exp (-i\om t)$ and integrating over $t$ from $0$ to $\infty$ leads to
\bea
\label{eq:J0}
\frac{1}{c_L^2}\frac{\partial J(\vec{k},0)}{\partial t} + k^2\xi J(\vec{k},0) + k^2 D(\vec{k},\omega) + i\frac{J(\vec{k},0)}{c_L^2}\omega &&  \\ 
- g^{-1}(\vec{k},\omega) {\cal J} (\vec k, \om) &=& 0 \nonumber 
\eea
where
\beq
D(\vec k,\om ) \equiv  \int_0^{\infty} dt e^{-i\om t} \langle j^*(\vec k,0) {\cal I} (\vec k, t) \rangle
\eeq
and
\beq
{\cal I} (\vec k, t) \equiv \int_0^{\infty} d \tau f(\tau +t) j(\vec k, -\tau ) 
\eeq
is a functional of the history of the system previous to its initial state $j(\vec{k},0)$. The initial value problem for dispersive waves, i.e., when the speed of propagation depends on frequency, does in general depend on its previous history. 
We shall assume there is no statistical correlation between the initial value of the current $j^*(\vec k,0)$ and its previous, dispersion-dominated, history:
\beq
D(\vec k,\om ) = 0.
\eeq
We now show that 
\beq
\frac{1}{c_L^2}\frac{\partial J(\vec{k},0)}{\partial t} + k^2\xi J(\vec{k},0)   = 0.
\label{eq:initzero}
\eeq
Indeed, note that Eqn. (\ref{eq:wek}) is the time derivative of 
\bea
    -\frac{1}{c_L^2}\frac{\partial j(\vec{k},t)}{\partial t} - k^2 j(\vec{k},t) - nk^2\xi{ u(\vec{k},t)} & & \nonumber \\
     + nk^2\int_{-\infty}^{\infty}dt'\ f(t') u(\vec{k},t-t') &=& 0 \, .
     \label{eq:wektime}
\eea
Since $\langle {\jmath}^*_{l \vec k}(0) u_{l\vec k} (0)\rangle =0$ because density and velocity are uncorrelated, and assuming again that the initial value of the current is uncorrelated with the previous, dispersion dominated, history, we have (\ref{eq:initzero}). Consequently, Eqns. (\ref{eq:J0}) and (\ref{eq:Jkw}) lead to
\beq
\label{eq:mainJg}
J(\vec k, \om) =-\frac{2}{c_L^2} J(k,0) \text{Im} \left[\om g(\vec k,\om) \right]
\eeq
{where $g(\vec k,\om)$ is the Green's function and }which leads to the density correlation through
\beq
\label{eq:denscorrel}
S(\vec k,\om ) = \frac{k^2}{\om^2} J(\vec k, \om) \, .
\eeq

\subsection{Relation of the BP to the strings}
\label{stringsnadbp}

Following \cite{Lund2015} we {describe the normal mode density of states of the model solid as}:
\beq
g(\omega) = g_D(\omega) + g_S(\omega)
\eeq
where $g_D(\om)$ is the Debye distribution and $g_S(\om)$ is the excess over this distribution (i.e., the boson peak). This distribution of excess modes over the Debye distribution determines a distribution of string lengths $L$ through
\beq
\label{eq:bpp(L)}
{g_S(\omega_0) d\omega_0 \equiv  2 p(L) dL }
\eeq
with $\om_0 = \sqrt{(\pi c /L)^2 -(B/2m)^2}$. {The factor of two in Eq. (\ref{eq:bpp(L)}) indicates that there are twice as many states of a given frequency as strings. This is due to the fact that each string can oscillate along two linearly independent directions that are perpendicular to its equilibrium position.}

{Here $c$ is the speed of waves along the string, $B/2m$ represents the internal damping of elastic waves in the string}.
So, given a specific Boson peak, as determined by atomistic modeling or experimental data, $g_S(\om_0)$ can be read off it and Eqn. (\ref{eq:bpp(L)}) determines the string length distribution $p(L)$, which in turn, through (\ref{eq:finalfom}), determines the index of refraction $f(\om)$ that goes into the Green's function (\ref{eq:fullgxi}) whose imaginary part provides the correlation function $S(\vec k, \om)$ using (\ref{eq:mainJg}) and (\ref{eq:denscorrel}).

\section{Relating the BP data to the IXS data}
\label{sec:exp} 
{In this section we relate the derived $S(\vec k, \om)$ to the experimental data that can be obtained in an inelastic x-ray  scattering experiment (IXS), in order to apply our model to the data collected with this technique on two prototype glasses, glycerol and silica. }

{The experimental inelastic scattering spectrum of a glass is composed of a Brillouin triplet: an elastic line, related to the frozen density fluctuations, and the Stokes and Anti-Stokes inelastic peaks corresponding to the annihilation and excitation of a vibrational excitation. The measured spectrum can be formally described as {\cite{PailhesPhonchap}}: 
\beq
\label{eq:iteo}
I(k,\omega) = A_k \hbar\omega\frac{[n_B(\omega)+1]}{k_BT}S_{tot}(k,\omega)
\eeq
}
{Here, we have replaced the vector $\vec k$ by its modulus, given the isotropy of the system. $A_k$ is a normalization factor reflecting the $k$ dependence of the atomic form factors,  $n_B(\omega)=(e^{\hbar\omega/k_BT}-1)^{-1}$ is the Bose-Einstein factor,
$S(k,\om)$ is the longitudinal density-density correlation function, which is modeled as the sum of a delta function for the elastic line and an inelastic contribution:}
\beq
\label{eq:stot}
S_{tot}(\vec k,\omega) = f_k\delta(\omega) + (1-f_k)S(k,\omega)
\eeq
{where $f_k$ accounts for the measured proportion between elastic and inelastic parts of spectra and is related to the non-ergodicity factor and $S(k,\omega)$ is given by (\ref{eq:denscorrel}). }

{Eq.~(\ref{eq:iteo}) can be fitted to the experimental spectra after convolution with the instrumental resolution function $R(\om)$, adding a $k$-dependent background $bg_k$:
\beq
\label{eq:iexp}
I_{exp}(k,\omega) = bg_k + \int{I(k,\om)R(\omega-\omega')d\omega'}
\eeq
}
Now that we have a reasonable expression for the intensity to work with, the next task is to assign physical values to the parameters of the model. 
For the instrumental resolution function, we use the one experimentally measured and we model it via a linear combination of a Lorentzian and a Gaussian function. We are then left with 3 groups of parameters for the description of the experimental spectra:

\begin{itemize}
\item Experimental setup and partly material-related: $bg_k$, $A_k$.
\item Material-related: $f_k$, $\xi$, $J(\vec k,0)$, $c_L$ and $c_T$.
\item Strings-related: $\alpha$, $c$, $B/m$ and $g_S(\omega)$.
\end{itemize}
The function $g_S(\omega)$ is obtained from the measured density of states, by modeling the excess over Debye distribution using a 5-degree polynomial ({\it i.e.} up to $\omega^5$). {Note that the theoretical $g(\om)$ is a number of states per unit frequency per unit volume. The experimental density of states is, however, normalized to the atomic volume, and thus is a number of states per unit frequency. The extra volume is obtained from the very low frequency asymptotics of the experimental values, where it does coincide with the Debye theory.}
The number of free parameters can be reduced by relating the velocities $c_L$, $c_T$ and $c$ through the equalities: $c_T = c_L/\sqrt{K/\mu+4/3}$, where $K$ and $\mu$ are the material's bulk and shear modulus respectively, and\cite{Lund1988} $c = c_T$. 
{The zero frequency value $J(k,0)$, is fixed to literature data of the longitudinal velocity as measured with Brillouin light scattering}. To simplify our model, we observe that the effect of damping in the elastic waves due to the internal friction of strings, given by $B/m$, can be reasonably assumed to be negligible compared to the other sources of damping. We thus take its value simply as zero.

{The remaining fitting parameters are thus $bg_k$, $A_k$ for the instrumental part, $f_k$, $\xi$ and $c_L$ for the material-part, and only $\alpha$ for the strings-related part. 
$\xi$, $c_L$ and $\alpha$ are expected, in a first approximation, to be $k$ independent. For this reason, and for getting the best description over a large $k$ range, they are found by simultaneously fitting 5 IXS spectra throughout the studied wavenumber range. The $k$ dependent parameters, such as  $bg_k$, $A_k$ and $f_k$ are then found by {independently fitting each spectrum using Eq.(\ref{eq:iexp}) }, for fixed values of $\xi$, $c_L$ and $\alpha$. The procedure is iteratively repeated until convergence.}

To sum up: given Boson peak data $g_S(\om)$, one obtains $p(L)$ through Eq. (\ref{eq:bpp(L)}). This relation involves a free parameter $B/2m$, the internal damping of string motion, that in the examples that will be worked out below will be put equal to zero. It also involves $c$, the speed of waves along the string, a parameter that is completely determined by {$c_T$, the transverse speed of sound} of the material under consideration. Next, $p(L)$ determines $f(\om)$ through Eq. (\ref{eq:fandp(L)}). This relation involves $m$, the mass per unit length of the string, which is related to the mass density $\rho$ of the material through $m = \alpha \rho \, b^2$, where $b$ is a length of order one inter-atomic distance that cancels out in the calculations, and $\alpha$ is a dimensionless parameter of order one that, in the examples below, will be effectively put equal to one. {Finally, as mentioned, $J(k,0)$ is also fixed from literature data}. With $f(\om)$ determined, $S(\vec k, \om)$ is found using Eqs. (\ref{eq:fullgxi}), (\ref{eq:mainJg}) and (\ref{eq:denscorrel}). This final formula for $S(\vec k, \om)$ depends on two parameters: $\xi$, the wave attenuation coefficient in the absence of strings, and $c_L$, the speed of sound. They will be varied, within reasonable bounds, to fit the $S(k,\om)$ obtained from the experimental $I(k,\om)$. This, in turn, is done using Eqs. (\ref{eq:iteo}) and (\ref{eq:stot}). These expressions depend on experimental parameters, $A_k$ and $bg_k$, and a material-related parameter, $f_k$.

Once the parameters that afford the best fit to  the experimental data are obtained, they can be used to obtain the phase velocity $v$ and attenuation $\Gamma$ solving for the poles, say $K(\om)$, of the Green's function in Eq.(\ref{eq:fullgxi}):
\bea
\label{phasevel}
v & = & \frac{\om}{\text{Re}[K(\om)]} \\
\Gamma & = & - c_L\text{Im} [K(\om)]
\label{attenuationGamma}
\eea

For testing the modeling of the inelastic scattering data through the string distribution obtained from the experimental density of states, we have chosen two prototype cases, a fragile glass, glycerol, and a strong one, silica, for which all the experimental data are available.

\subsection{The case of glycerol}
\label{sec:glycerol}
We present here the fit of inelastic x ray scattering data obtained by Monaco and Giordano~\cite{Monaco2009} on a glass of glycerol at 150~K, the glass transition temperature being $T_g=189$~K. 
In that paper, data between $k=1.1$ and $k=5$~nm$^{-1}$ have been reported and fitted with a damped harmonic oscillator (DHO) model for the inelastic contribution. Such work has given the first experimental evidence of the existence of elastic anomalies in glasses in the Boson Peak region and the early breakdown of the Debye approximation: a crossover from a regime where the phonon attenuation increases with $k^4$ to a regime where it goes as $k^2$, accompanied by a dip in the sound velocity with respect to the expected behavior. Moreover, the calculation of the density of states using the experimental $S(k,\om)$ has been found to be in  reasonable agreement with the experimental data as measured by inelastic neutron scattering at 170~K~\cite{Wuttke1995}. The role of the attenuation has been there highlighted, as fundamental to get the right position of the Boson Peak.

\begin{figure}[h!]
\includegraphics[width=8cm]{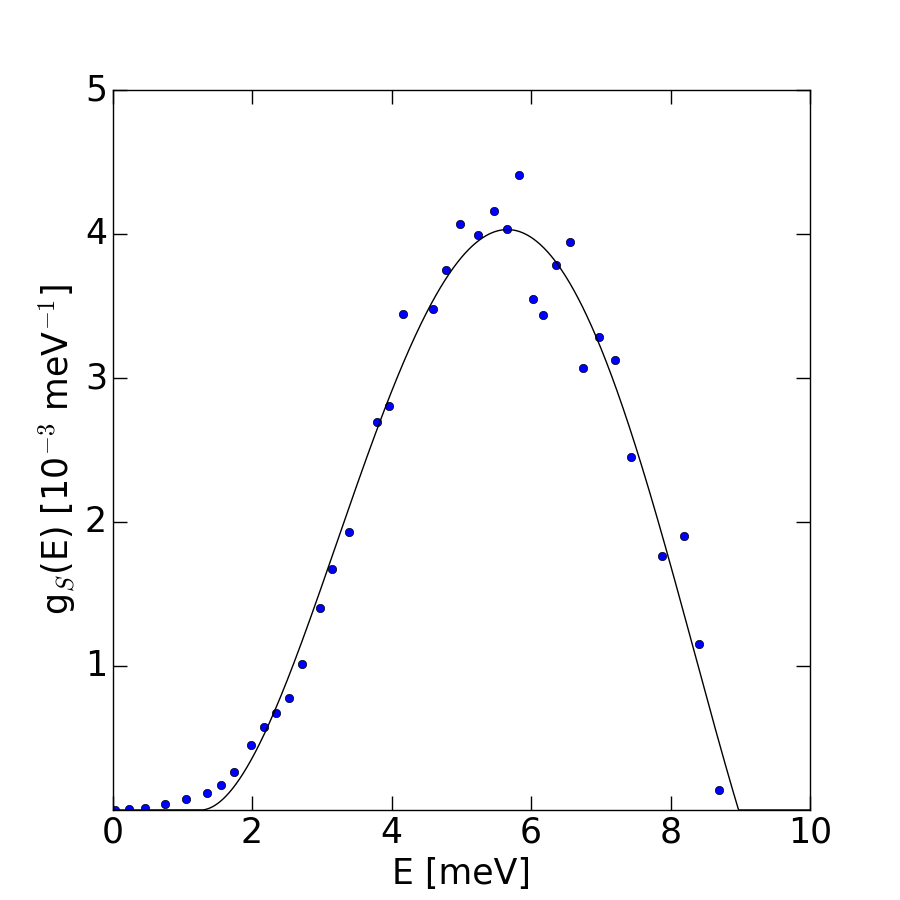}
\caption{The Boson Peak of glassy glycerol at 170~K, {taken from Ref.~\cite{Wuttke1995}, and reported in non-reduced units ($g(\omega)-g_D(\omega)$), together with the fit with a 5 degree polynomial}}
\label{glyBP}
\end{figure}

\begin{figure}[h!]
\includegraphics[width=8cm]{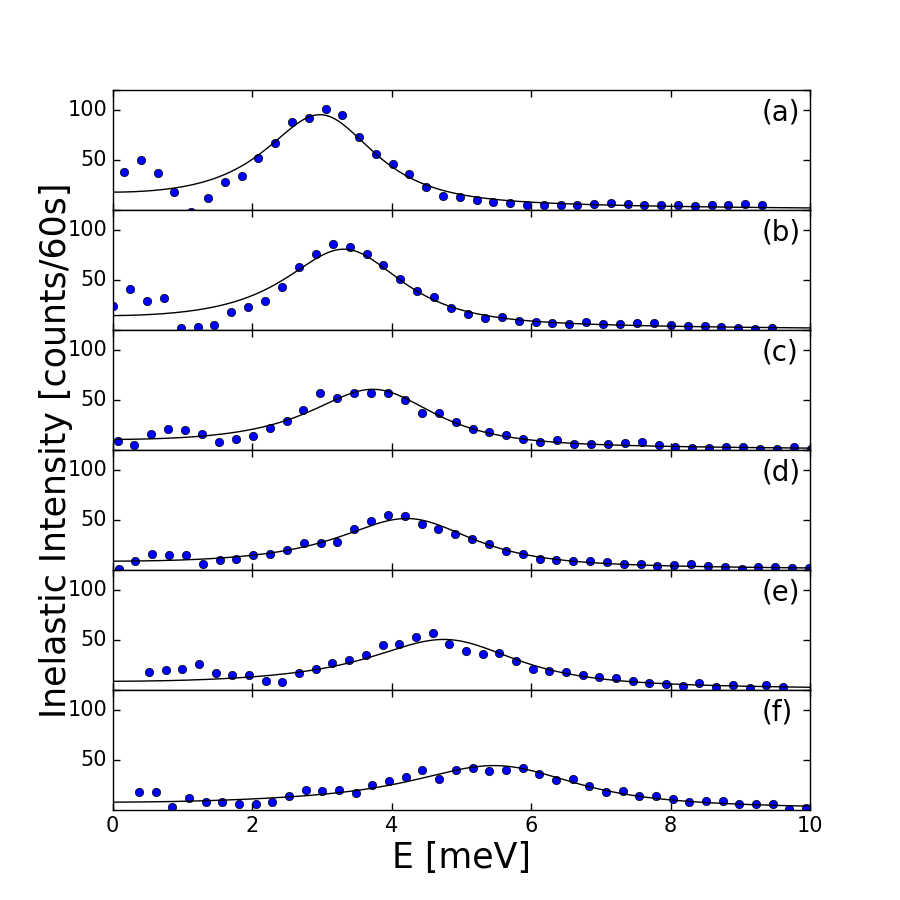}
\caption{Inelastic x ray scattering spectra of glassy glycerol as measured at 150~K and reported in Ref.~\cite{Monaco2009} for en exchanged momentum $q=$ 1.3, 1.5, 1.7, 1.9, 2.1 and 2.5~nm$^{-1}$ from panel a to f. For the sake of clarity we report only the positive energy exchange side and only the inelastic contribution after subtraction of the elastic line. The {black solid} lines represent the fit with the string model, where the string distribution has been deduced from the Boson Peak.}
\label{glyIXS}
\end{figure}

If in that work the authors could reproduce the density of states from the experimental density-density correlation function, here we follow the opposite direction, using the density of states for reproducing the $S(k,\om)$. 
Figure~\ref{glyBP}, reports the excess density of states on the Debye prediction as obtained from Ref.~\cite{Wuttke1995} and its fit with a 5-degree polynomial, which will be used for the successive fit of the inelastic data. 
Figure~\ref{glyIXS} reports a selection of inelastic spectra of glycerol, in the [1-3]~nm$^{-1}$ range, together with the fit to Eq.~\ref{eq:iteo}. For a better visualization, we have chosen to report only the inelastic part of the spectra, after subtraction of the fitted elastic line, and only for positive exchanged energies. 

We remind here that in Ref.~\cite{Monaco2009} the $k^4-k^2$ crossover was identified at $k\approx2$~nm$^{-1}$. The $k$ values here chosen span then the interesting $k$ region. We have not reported larger $k$ as the vibrational modes are largely too damped and the energy range too short for a constraining fit with our model. From the figure, it is clear that the fitting model catches quite well the $k$ dependence of the major inelastic contribution features (energy position and attenuation), with a fitting quality comparable to the one reported in Ref.~\cite{Monaco2009}. {The $k$-independent parameters fitted with our iterative procedure are reported in table \ref{tabfit}.}

\begin{figure}[h!]
\includegraphics[width=8cm]{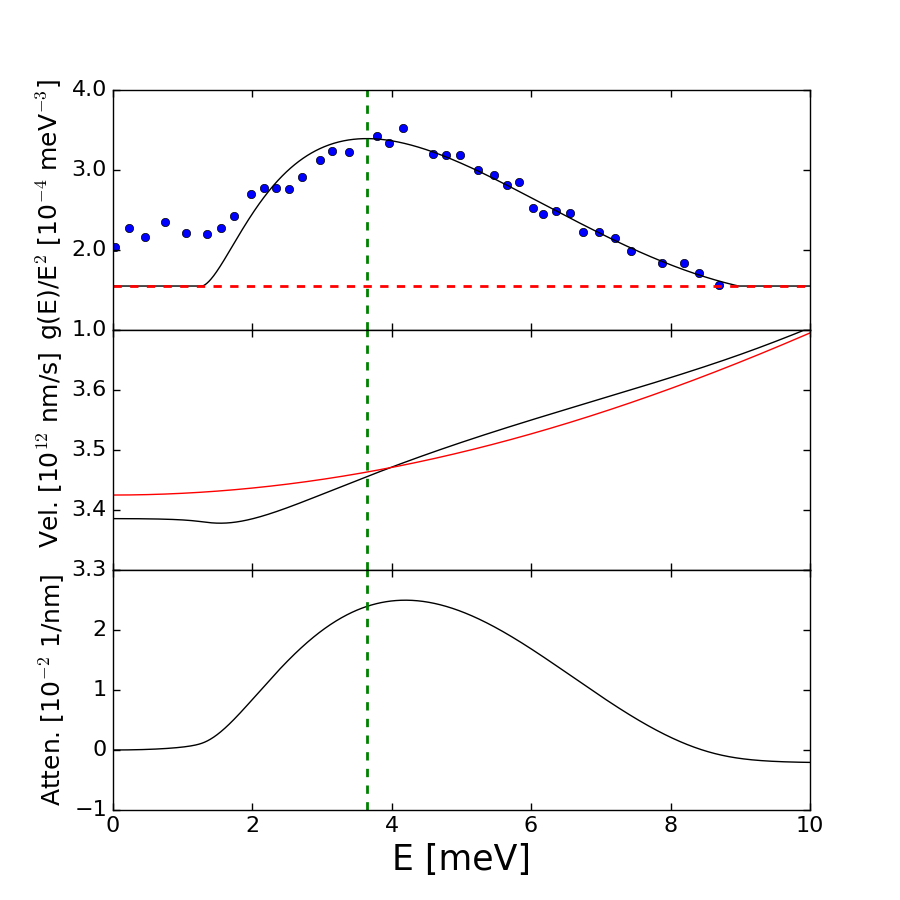}
\caption{Top panel: excess of modes (Boson Peak) in the regular density of states of glassy glycerol and its modeling with a 5-degree polynomial. The red dashed line represents what we should have in the absence of strings (no excess). The mismatch at low energy is due to the larger weight  that low-energy states have in this case, compared with the fit in non-reduced units reported on Figure \ref{glyBP}. Central panel: the velocity as obtained using Eq. (\ref{phasevel}), compared with the one in the case of no strings (red). Bottom panel: the excess attenuation, due to strings, with respect to the attenuation in the case of no strings. Its maximum, marked by the vertical green dashed line, corresponds to the change of behavior in the velocity from a lower to a higher velocity with respect to the no-string case, and to the maximum of the BP.}
\label{velattgly}
\end{figure}

{To go further, we can extract from our model the phase velocity $v$ and the attenuation $\Gamma$ calculated as indicated in Eqs (\ref{phasevel}-\ref{attenuationGamma}). They are reported in Figure \ref{velattgly}, together with the Boson Peak, and they are compared with the expected velocity and attenuation in absence of strings (in red).  It is clear that the addition of strings strongly modifies velocity and attenuation. More precisely, from these graphs, we understand that the longest strings, corresponding to the low energy modes of the BP, and thus to its onset, interfere with elastic waves slowing them down, while the shortest strings, corresponding to the high energy part of the BP, speed them up again. This interference is accompanied by the onset of a strong attenuation. Note that in the figure we report the ``string-dependent'' attenuation, {\it i.e.} the residual attenuation once the one without strings has been subtracted. As such, at low energy it is zero and constant, meaning that the total attenuation goes like $k^2$, and then it increases, meaning that the total one goes like $k^4$, before decreasing again. The maximum corresponds to the maximum of the BP, {\it i.e.}  the maximum of density of string modes. It is worth noticing that, while our model well reproduces the existence of a dip in the sound velocity as well as the $k^4-k^2$ crossover, all the phenomenology appears shifted at lower $k$ with respect to what is found with a DHO model. We don't expect the two models to give exactly the same results, as we explain in Appendix \ref{sec:DHO}, still the physical meaning behind the results obtained with the two models is very similar: we have the same elastic anomalies, which occur in both cases in correspondence with the BP, excess of modes in the regular density of states (string case) or the reduced one (divided by $\omega^2$, DHO case).}

\begin{table}
\begin{tabular}{|c|c|c|}
\hline
 & glycerol & silica\\
\hline
\hline
$J(k,0) (Km/s)$ & 3.625\cite{Comez2003} & 6.5\cite{Baldi2010}\\
\hline
$K$ (GPa) & 14.1917\cite{Scarponi2004,Lyapin2017} & 44.97\cite{Pabst2013}\\
\hline
$\mu$ (GPa) & 4.85\cite{Scarponi2004,Lyapin2017}  & 36.07\cite{Pabst2013}\\
\hline
$c_L$ (Km/s) &  {3.42$\pm$ 0.03} & {6.08 $\pm$ 0.05}\\
\hline
$\xi$ ($10^{-14}$ s) & {3.1 $\pm$ 0.3} &  {2.84 $\pm$  0.17} \\
\hline
\end{tabular}
\caption{Table of the physical quantities needed for the string model and the $k$-independent fitting parameters for glycerol and silica. }
\label{tabfit}
\end{table}

\subsection{The case of silica}
\label{sec:silica}
\begin{figure}[h!]
\includegraphics[width=8cm]{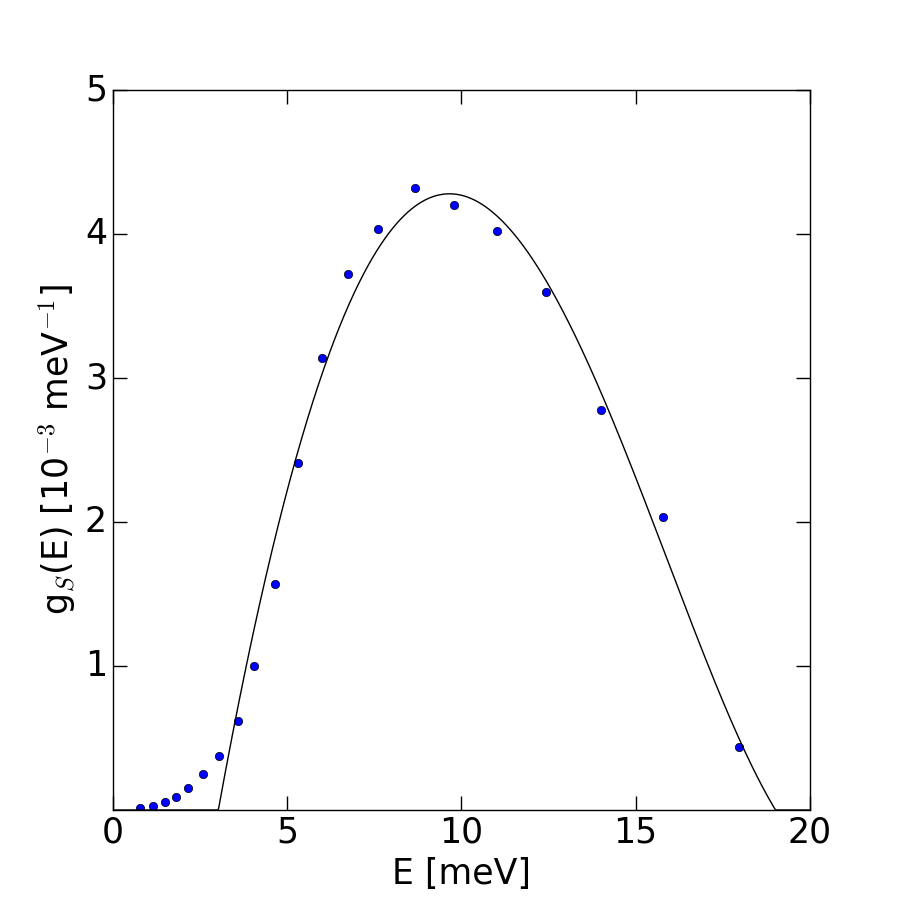}
\caption{The Boson Peak of glassy silica at 1673~K, { taken from Ref.~\cite{Baldi2011}, reported in non reduced units ($g(\omega)-g_D(\omega)$), together with the fit with a 5 degree polynomial}}
\label{silicaBP}
\end{figure}

\begin{figure}[h!]
\includegraphics[width=8cm]{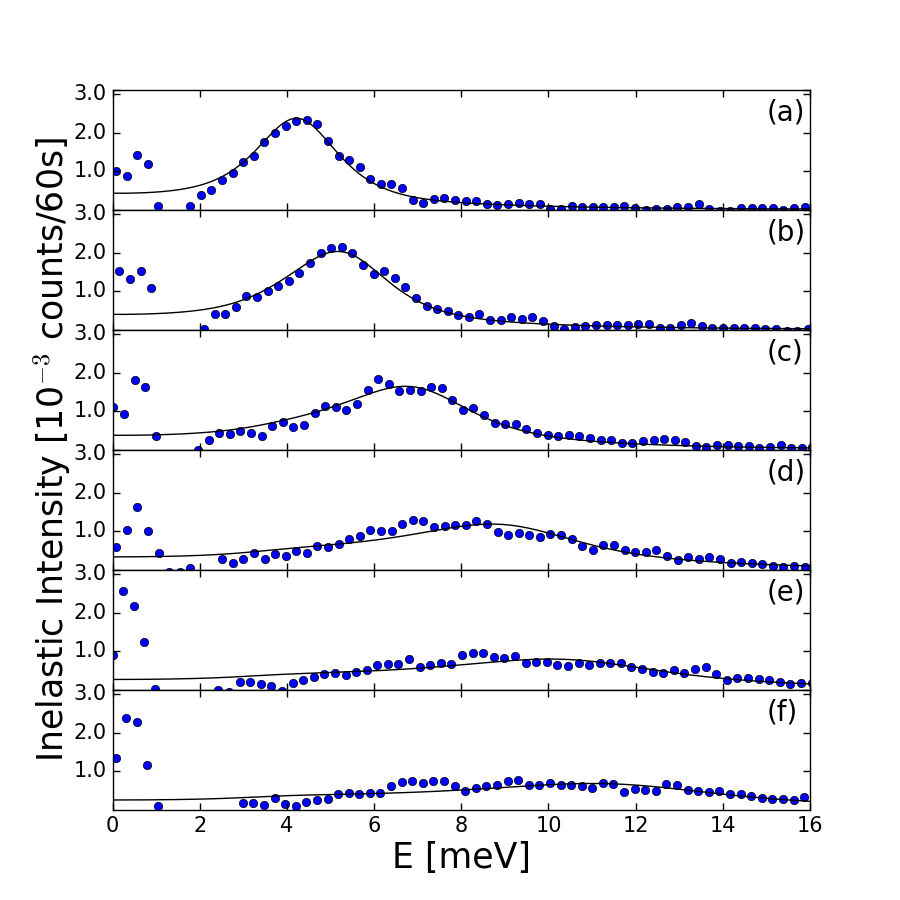}
\caption{Inelastic x ray scattering spectra of glassy silica as measured at 1620~K and reported in Ref.~\cite{Baldi2011} for en exchanged momentum $q=$1.1, 1.3, 1.7, 2.3, 2.7 and 2.9~nm$^{-1}$  from panel a to f. For the sake of clarity we report only the positive energy exchange side and only the inelastic contribution after subtraction of the elastic line. The {black} lines represent the fit with the string model, where the string distribution has been deduced from the Boson Peak.}
\label{silicaIXS}
\end{figure}

In this section, we apply the string model to the case of silica and more specifically we fit the data collected by x ray inelastic scattering at $T = 1620$~K {(Glass transition temperature $T_g=1450$~K\cite{Baldi2008})} and reported by Baldi et al.~\cite{Baldi2011}. Figure~\ref{silicaBP} reports the {excess density of states in non-reduced units as obtained from the one }measured by neutron scattering at $T=1673$~K and reported in the same paper, together with a 5-order polynomial fit for extracting the string distribution.  
Some selected inelastic spectra with the fit performed with our model in the exchanged wavevector range between 0.95 and 2.8~nm$^{-1}$ are reported in Figure~\ref{silicaIXS}. The agreement is quite good, up to $k \sim 2$~nm$^{-1}$, from which value, it gets worse and worse with increasing wavevector, overestimating the position and underestimating the width. {$k$-independent fitting parameters are reported in table \ref{tabfit}.}
{This can be understood going back to Ref~\cite{Baldi2011}, where a critical wavevector $k_c=1.41$~nm$^{-1}$ was identified, such that $\Gamma \propto k^4$ for $k \ll k_c$ and $\Gamma \propto k^2$ above. Related to the Ioffe-Regel crossover, this critical wavevector has been successively identified as the limit above which the spectra are mostly determined by the local order and similar to the ones of the polycrystal with the same local order and density. Here the peak acquires contributions from the transverse modes and from the higher energy optic modes, not being then a single plane wave anymore~\cite{Baldi2013}. We can thus expect our model not to work anymore in this regime.}

\begin{figure}[h!]
\includegraphics[width=8cm]{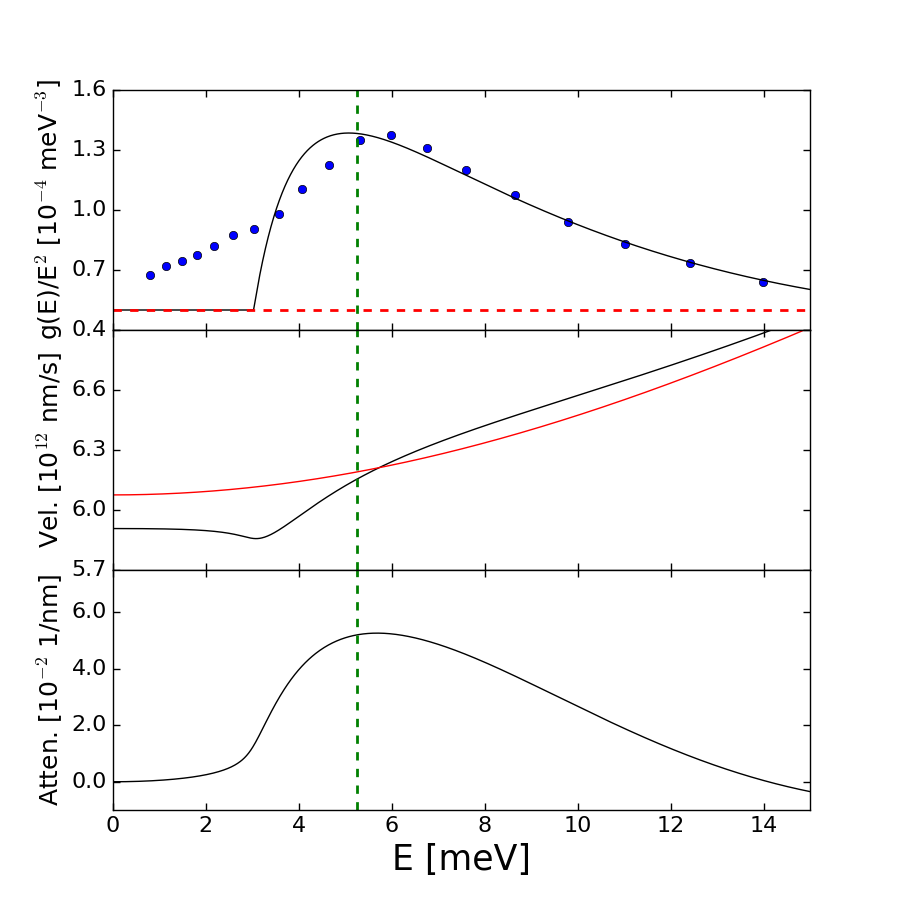}
\caption{Top panel: excess of modes (Boson Peak) in the regular density of states of glassy silica and its modeling with a 5-degree polynomial. As in glycerol, the mismatch at low energy is due to the larger weight  that low-energy states have in this case, compared with the fit in non-reduced units reported on Figure \ref{silicaBP}. The red dashed line represents what we should have in the absence of strings (no excess). Central panel: the velocity as obtained using Eq.\ref{phasevel}, compared with the one in the case of no strings (red). Bottom panel: the excess attenuation, due to strings, with respect to the attenuation in the case of no strings. Its maximum, marked by the vertical green dashed line, corresponds to the change of behavior in the velocity from a lower to a higher velocity with respect to the no-string case, and to the maximum of the BP.}
\label{velattsil}
\end{figure}

{Similarly to the case of glycerol, we report in Figure \ref{velattsil} the Boson Peak, the velocity as modified by the presence of strings, and the excess attenuation due to the strings. The same considerations as above hold true for silica, both on the effect of strings and on the different crossover position with respect to the findings of the DHO model.}

\section{Discussion}
\label{sec:discussion}
\subsection{The string distribution}
{We have seen that our string model is able to provide a fit to experimental data as good as the one historically obtained with the DHO model. The derived velocity and attenuation, as calculated from the fitting parameters and Eqs. (\ref{phasevel}) and (\ref{attenuationGamma}), can be clearly interpreted in terms of interference of the elastic waves with soft localized modes from strings with different lengths, giving a straightforward understanding of the elastic anomalies in glasses. }

\begin{figure}[h!]
\includegraphics[width=8cm]{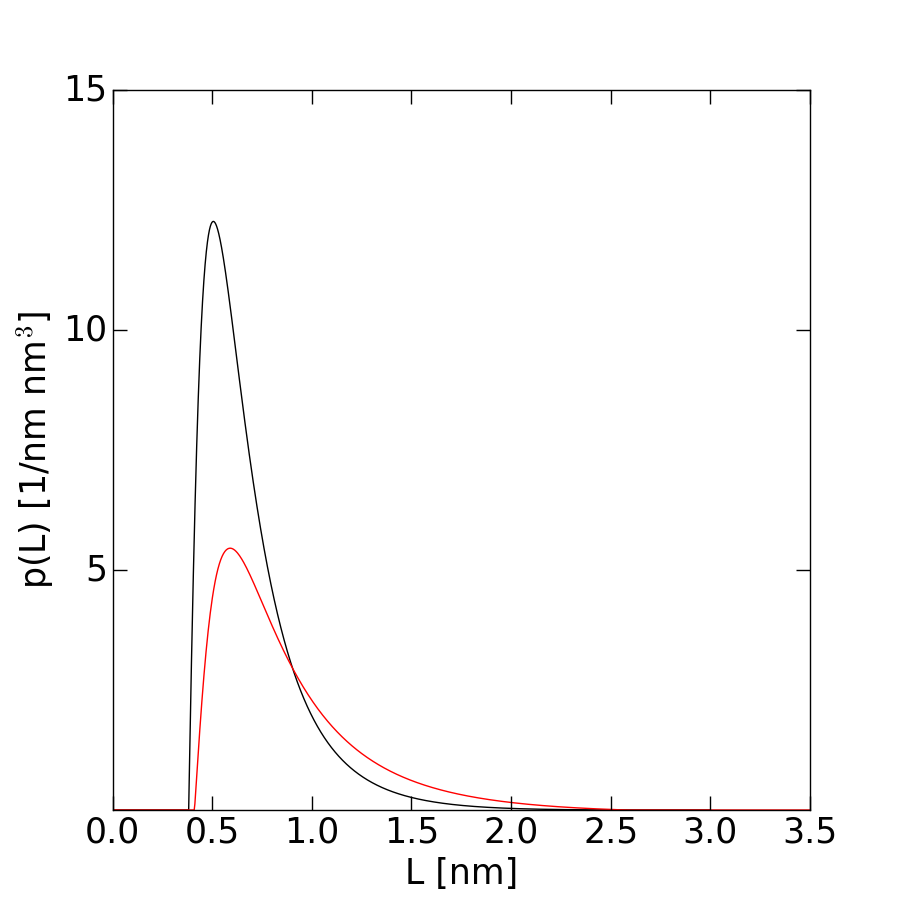}
\caption{We compare here the string length distribution as obtained from the Boson Peak in glassy glycerol (black) and in silica (red). }
\label{Ldistr}
\end{figure}

{It is interesting to underline that in the examples above we have made a series of simplifications, which allows to extract the important parameters of the model: i) we have neglected the intrinsic damping in the strings, by imposing $B/m=0$; {ii) we have assumed $\alpha=1$, which corresponds to say that the linear mass density of the string is the same as in the material and (iii) we have taken the speed of sound $c$ in the string equal to the speed of shear waves: $c=c_T$. Moreover, the excess density of states is fixed by the experimental observation.}
The parameter $b$ (cross-sectional size of the string), still present in the model, cancels out, so that the dynamics depends only on the strings length, whose distribution leads to the distribution of string vibrational modes and then to the Boson Peak.
As such, our fitting model is almost with no free parameters once the Boson Peak has been measured, which univocally fixes the string length distribution. The quality of the agreement of the model with the data is then impressive. In Fig.\ref{Ldistr} we report the length distribution obtained for glycerol and silica. }

{Despite the large structural and dynamical differences between these two glasses, the Boson Peak is located at very similar energies, with a maximum, in reduced units, at 4 meV  in glycerol and 6 meV in silica. This may explain the similarity of the string length distributions reported in the figure, with a slight difference in the position of the maximum which well mimics the one in the BP position, being the maximum 0.51 nm and 0.6 nm for glycerol and silica respectively. Such value corresponds well to the size of the elastic heterogeneities as found for silica in \cite{Baldi2013}, and is also the typical size of the dynamical heterogeneities in glass forming liquids close to the glass transition, suggesting a direct link between our strings and these latter \cite{Tracht1998,Berthier2005}.
It is also remarkable that in both glasses the distribution spans from about 0.4 nm up to 2.6 nm, implying that the same string modes are present in both glasses. Moreover, the beginning of the distribution, at 0.4nm,  corresponds well to the correlation length for the spatial fluctuations of the local sound velocity, as identified in silica in \cite{Baldi2014}.} 
Beyond these similarities, the two glasses distributions differ for their integral area, which is directly connected to the Boson Peak amplitude, and corresponds to the total string number density: {{$4.678$ nm$^{-3}$ for glycerol and $3.129$ nm$^{-3}$ for silica}. 
In light of these observations, one may wonder if the extent of the string distribution as well as a maximum around 0.5 nm are somehow universal parameters independent on the glass and on the position and intensity of the BP, which are only dependent on the finer details of the distribution (such as the precise position of the maximum and the total integral).
Glycerol being an intermediate glass, it would be interesting to apply this model to a very fragile glass to shed light on this point. More generally, applications of the model to more systems would help in understanding the role played by the length distribution detailed shape in determining the excess of modes position and shape (the BP), and its correlation with the fragility of the glass.  }

\subsection{Strings and soft localized modes from numerical simulations}
{Very recently large numerical simulations have proved the existence in glasses of soft localized non-phononic modes at low energy. It is interesting to investigate further the link which can exist between the strings in our theory and such modes.} These latter correspond to particle displacements decaying as $r^{-2}$ in 3D from the localization point~\cite{Shimada2018,Lerner2016,Gartner2016,Kapeteijns2018,Lerner2018}, the same decay law being obeyed by the plastic events associated with quasi-static deformations. Interestingly, their density of states grows like frequency to the fourth in two, three, as well as four dimensions for energies below the Boson Peak~\cite{Lerner2018}.  
In order to check wether the string vibrations present the same characteristics as the soft localized modes reported in these works, we calculate the particle displacement associated with them, solving Eqs. (\ref{EQondeMaurel2005}-\ref{termesource}).

\begin{figure}
\includegraphics[width=0.4\columnwidth]{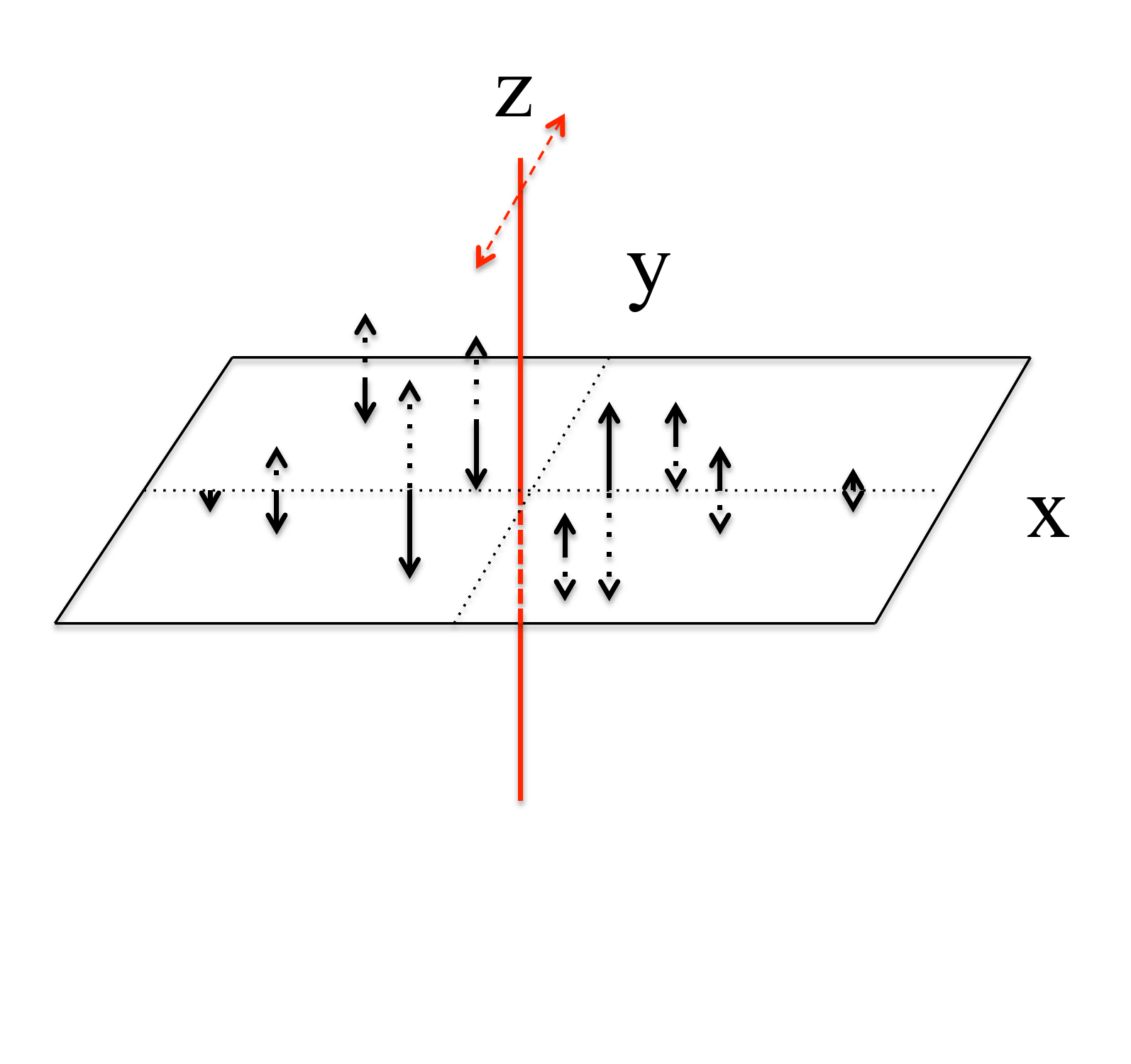}
\includegraphics[width=0.4\columnwidth]{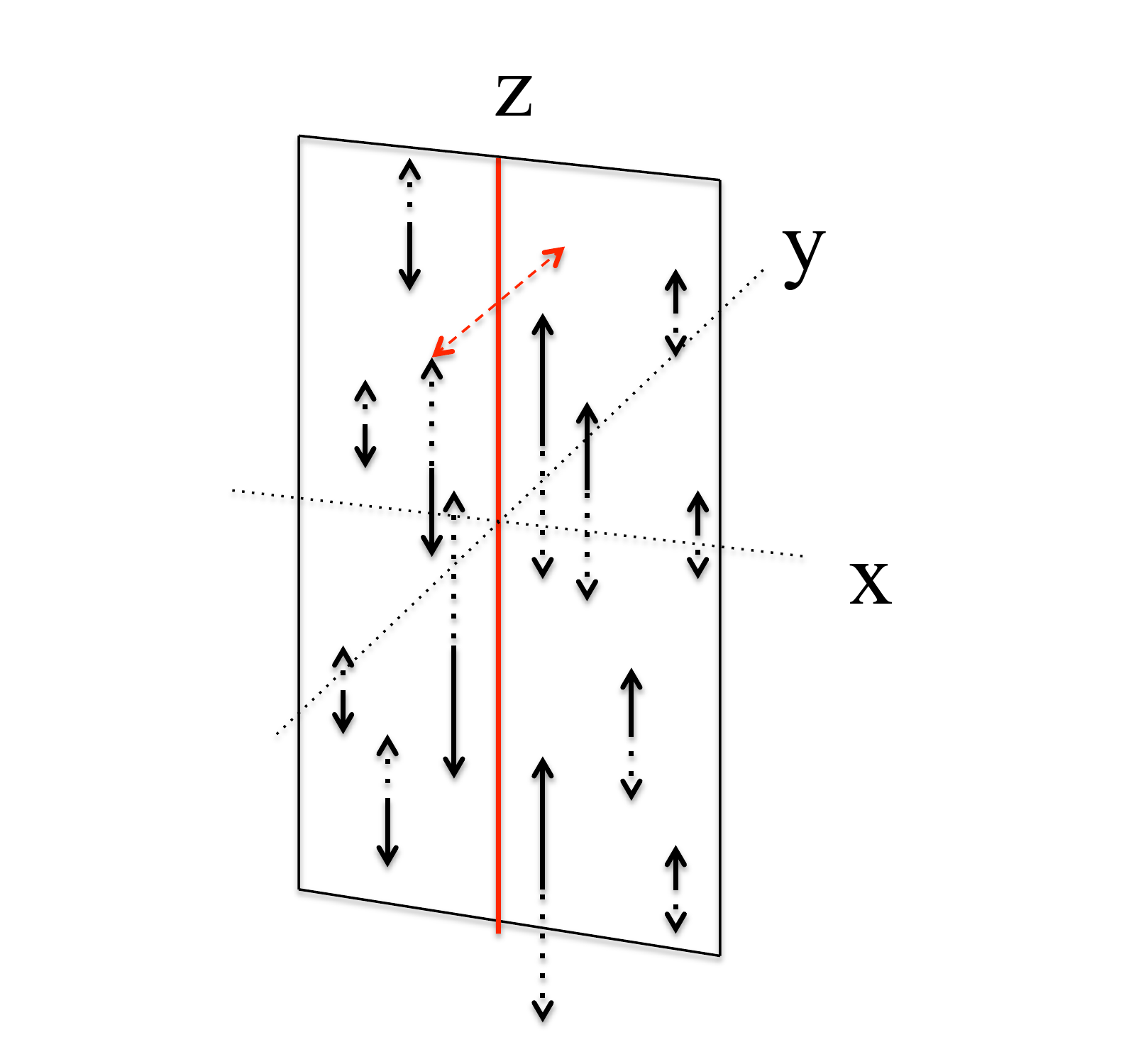}
\caption{Particle motion associated with string motion, according to Eqs. (\ref{stringy}) and (\ref{u3motion}), for particles very close to the string ($r \ll L$). The string (red straight line) points along the z-axis and moves perpendicular to itself, along the y-axis. Particles  move along a direction parallel to the string, the z-axis. Left panel: motion of particles lying along the x-y plane. Right panel: motion of particles along x-z. Here the particle motion is invariant with respect to translations along the z-axis. Note that the string and its (perpendicular) motion determine a plane, here the $y-z$ plane. Particle motion on one side of this plane, say positive $x$, is 180$^{\circ}$ out of phase with respect to particle motion on the other side, negative $x$.}
\label{sepulveda}
\end{figure}

{Consider a string segment of length $L$ along the $z$-axis, oscillating with a small amplitude $\delta$ and fundamental frequency $\om_0$ along the $y$-direction. {The displacement $X_n(s,t')$ as a function of a Lagrangian variable $s$ along the string and time $t'$, is }
\beq
X_n(s,t') = (0, \delta e^{i\om_0 t'} \sin \left[ \frac{\pi}{L} (s+\frac L2 ) \right],s)
\label{stringy}
\eeq
with $-L/2 <s<L/2$, equilibrium position $X^0_n (s) = (0,0,s)$, $b_i =(0,0,1)$ {where $m=\rho b^2$} and $\tau_j = (0,0,1)$ is the tangent to the string (see Appendix \ref{Apa}). In this case the particle velocity obtained from (\ref{EQondeMaurel2005})-(\ref{termesource}) is
\begin{widetext}
\beq
\label{Muravelscrew}
v_m(\vec x, \om_0) = -i b \mu \om_0 \delta \int_{-L/2}^{L/2} ds \sin \left[ \frac{\pi}{L} (s+\frac L2 ) \right] \left( \nabla_1 G^0_{3m} (\vec x - \vec X_0(s), \om_0) + \nabla_3 G^0_{1m} (\vec x - \vec X_0(s), \om_0) \right)
\eeq
where $G^0_{ij}$ is the bare Green's function introduced in Appendix \ref{Apa}. For our present purposes we need it in the representation \cite{Love_book}
\bea
4\pi \rho G^0_{km}(\vec r, \om_0)  &=&  \frac{1}{c_T^2 r} \delta_{km}  e^{-i \om_0 r/c_T} + \frac{r_k r_m}{r^3} \left[ \frac{1}{c_L^2} e^{-i \om_0 r/c_L} -  \frac{1}{c_T^2} e^{-i \om_0 r/c_T} \right]  \nonumber \\ 
 &&\hspace{2em}  + \left( \frac 1r \right)_{,km} \left[ \frac{-1}{i\om_0} \left(  \frac{r}{ c_T} e^{-i \om_0 r/c_T} - \frac{r}{ c_L} e^{-i \om_0 r/c_L} \right)+ \frac{1}{\om_0^2} \left(  e^{-i \om_0 r/c_T} -  e^{-i \om_0 r/c_L} \right) \right] \, .
\eea
\end{widetext}
After substitution of this expression into (\ref{Muravelscrew}) {we can identify three distinct regimes} as a function of distance from the string: 
\begin{itemize}
\item[a)] $r\ll L$; that is, particle motion very close to the string, at distances small compared to string length. In this case it is straightforward to verify that, far from the pinning points, $v_1=v_2 =0$ and that 
\beq
u_3 (x,y;\om_0) \propto \frac{x}{x^2+y^2}
\label{u3motion}
\eeq
where $(x,y)$ are coordinates on a plane centered on the string and perpendicular to it. This means that particles sharing the same $(x,y)$ coordinates will have the same motion, parallel to the string. {This latter however moves perpendicular to itself, as it moves along the $y$ direction while it is aligned along the $z$ direction. As such, the particle displacement generated by the string corresponds to a transverse excitation, as can be better appreciated from Fig.~\ref{sepulveda}}. This motion seems compatible with what reported by Schober et al. \cite{Schober1993,Schober1997,Schober2002} in numerical simulations.
\item[b)] $r \sim L$; distances on the order of the string length. Remember that $\om_0 \sim c/L$. In this case, omitting terms depending on the string acceleration, which will be considered in (c), the particle displacement decays like $r^{-2}$ away from the center of the string, in a very nice agreement with the results of Lerner et al. \cite{Lerner2016,Gartner2016,Kapeteijns2018,Lerner2018} for the eigenvectors of the localized modes of a disordered solid. 
\item[c)] $r \gg L$; distances very far from the string. In this case $v_m (r, \om_0) \sim 1/r$. This is the typical decay of a radiation field and was to be expected. Indeed, the vibrating string picture encapsulates the time variation of an inhomogeneous strain distribution. An inevitable result of such a time evolution is the generation of a radiating field that propagates away to infinity. This part of the particle displacement however emerges only after considering its time evolution, it depends on the string acceleration and can not be obtained from a diagonalization of the dynamical matrix as calculated in Refs. \cite{Shimada2018,Lerner2016,Gartner2016,Kapeteijns2018,Lerner2018} 
\end{itemize}
So the string plays the role of a one-dimensional dynamical object that encapsulates the coherent motion of a localized, three-dimensional, set of particles.
}

{Once we have assessed that the particle displacement related to a string motion is compatible with the one of the non-phonon soft localized modes, we next need to understand if the density of states of the localized modes is also the same. The low energy $\sim \om^4$ behavior is actually not incompatible with the experimental density of states we have used in previous sections. In terms of strings, it would translate into a density of string lengths $\sim L^{-6}$ at large L.}

{Finally, the strings we have been considering are a specific implementation of non-affine displacements. As such, they can account both for the excess number of vibrational states, as well as the dispersive properties of acoustic waves, in glasses. In this sense, our work is intimately in agreement with the recent results by Caroli and Lema\^itre, who have directly solved the Newton's equation for a system of particles interacting through a soft-sphere potential in two dimensions, and highlighted the major role played by non-affine displacements in the acoustic behavior of amorphous solids~\cite{Caroli2019}. It would be interesting to extend the string picture to two dimensional glasses to further relate this picture to the cited findings. }

\section{Conclusion}
\label{sec:conclusion}

{We have presented in this paper an implementation of the analytical string theory as developed by one of us for the description of the physics of dislocations, and later applied to the description of glasses, modeled as continuum media with embedded elastic strings heterogeneities\cite{Lund2015}.}
{In that previous work it was already shown that such a description can account for the elastic anomalies reported at frequencies comparable with the Boson Peak: the strong phonon scattering and the negative dispersion in the sound velocity. Here we further develop that theory, in order to get the density density correlation function and compare with experimental data obtained by inelastic x ray scattering on two prototypical glasses, glycerol and silica. }

{We show that, once the vibrational density of states has been measured, we can use it for univocally fixing the string length distribution inherent to that glassy system. The density density correlation function obtained using such distribution is strongly constrained, and able to account for the experimental data in a fairly extended wavevector range, with the refinement of just a few fitting parameters.
As such, the quality of the fit in the large wavevector range around the Boson Peak region is quite impressive for both systems.} 

{From such fits, we can infer that the sound velocity and the attenuation of the elastic waves traveling in the continuum medium are perturbed due to the interference with the string vibrational eigenmodes: longer strings, characterized by low energy eigenmodes, slow down the sound wave, while shorter strings, whose eigenmodes correspond to the high frequency tail of the Boson peak, speed it up again. The maximum of the BP corresponds to a critical energy at which the change from negative to positive interference in the speed of sound takes place, as well as to the maximum of the extra-attenuation, due to the interference with the strings. While the phenomenology closely resembles the one evidenced in\cite{Monaco2009}, this critical energy is slightly smaller. However, as explained in Appendix \ref{sec:DHO}, it is not expected to obtain the same value. }

{We have shown that the string distribution model is compatible with the recent theoretical advances in the understanding of the vibrational properties of glasses in the subTHz-THz range. More specifically, the atomic motion associated to the strings presents a similar spacial amplitude behavior as the one obtained for the low energy non-phononic modes in recent numerical simulations, allowing to directly connect such non-phononic soft excess modes with the fundamental eigenmodes of the elastic heterogeneities represented by the strings.}

{It is thus interesting to critically inspect the string length distribution and its connection with the vibrational density of states and elastic anomalies. We have evidenced some common features, such as the range of string lengths and a peak at {about 0.5} nm, which are compatible with the reported elastic heterogeneities correlation length for silica\cite{Baldi2013} and dynamical heterogeneities in supercooled liquids\cite{Tracht1998,Berthier2005}. 
Such common properties suggest a possible universality, beyond the fragility of the glass,{ which seems to be more closely related to the string density .} }

{Our work thus not only offers a powerful tool, allowing to directly obtain the longitudinal density density correlation function from the density of states measurement, but also offers a microscopic understanding of the vibrational properties of glasses in terms of interference of the sound plane waves with soft modes of elastic 1D heterogeneities, compatible with the most recent theoretical findings. Such a picture allows to identify the string length distribution features as key players in determining the vibrational anomalies, evidencing the universal character of some of them against the material specific character of others.  For shedding light onto the relationship between string length distribution and fragility or connectivity, a systematic investigation of glasses with very different fragility and connectivity should be performed.}

\begin{acknowledgments}
This work has been supported by Fondecyt Grant 1191179 and ECOS-CONICYT Grant C17E02. 
\end{acknowledgments}

\begin{appendix}
\section{Coherent waves, strings, and the Boson peak \cite{Lund2015,Churochkin2016,Mujica2012}}
\label{Apa}
Consider a set of elastic strings of length $L$ with pinned end points. There are $p(L)dL$ strings with length between $L$ and $L+dL$ per unit volume. Each string is described by a displacement $\vec X(s,t)$ away from a straight equilibrium position $\vec X_0$, with $s$ a Lagrangian parameter along the string, that obeys the equation of an elastic string loaded by an external stress which, in the present context, will be associated with an elastic wave:
\beq
\label{screwstring}
m {\ddot{X}}_k( s ,t)+ B{\dot{X}}_k( s ,t)- {\cal T} X_k''( s ,t) =   \mu b \; {\mathsf N}_{kjp}\nabla_j u_p( \vec X ,t) 
\eeq
where $m \sim \rho b^2$ is a mass per unit length, $b$ is a length of order one inter-atomic distance, ${\cal T} \sim \mu b^2$ is a line tension, $B$ is a phenomenological damping coefficient, $\nabla_l u_k( {\vec X}_0 ,t)$ is the gradient of the displacement $\vec u (\vec x,t)$ associated with the incoming wave, evaluated at the string equilibrium position. ${\mathsf N}_{kjp} \equiv \epsilon_{kjm} \tau_m \tau_p + \epsilon_{kpm} \tau_m \tau_j$, where $\tau_j$ is the unit tangent along the string equilibrium position. A prime means differentiation with respect to $s$. 
The right-hand-side of (\ref{screwstring}) is the Peach-Koehler force for 
screw dislocations in elastic continua. This coupling ensures that only the shear modulus, and not the bulk modulus, will, later on, become frequency dependent, a fact that has been observed in the numerical simulations of Maruzzo et al. \cite{Maruzzo2013}.

On the other hand, an elastic wave travelling in the presence of such a string will obey the equation\cite{Maurel2005a,Maurel2005b}
\begin{equation}
\rho \ddot v_i(\vec x,t)-c_{ijkl}\nabla_j \nabla_l v_k
(\vec x,t)=s_i(\vec x,t),
\label{EQondeMaurel2005}
\end{equation}
with
\begin{equation}
 s_i(\vec x,t)=c_{ijkl}\epsilon_{mnk} b \int_{\cal L}
d s \; \dot X_m( s ,t) \tau_n \tau_l \nabla_j  \delta ( x-X_0 )
\label{termesource}
\end{equation} 
where $c_{ijkl} = \lambda \delta_{ij} \delta_{kl} + \mu (\delta_{ik}\delta_{jl} +  \delta_{il}\delta_{jk})$.

So the picture is as follows: an elastic wave hits an elastic string; the string responds according to Eqn. (\ref{screwstring}); this response is plugged into the right hand side of Eqn. (\ref{EQondeMaurel2005}) to obtain the scattered field generated by this response. Actually plugging the solution of  (\ref{screwstring}) into (\ref{EQondeMaurel2005}) leads to, in the frequency domain,
\begin{equation}
-\rho \omega^2v_i(\vec x,\omega)-c_{ijkl}\nabla_j \nabla_l v_k
(\vec x,\omega)= V_{ik} v_k (\vec x,\omega)
\label{EQondesV}
\end{equation}
where
\begin{equation}
\left. V_{ik}=  {\cal A} \; \mathsf N_{mij}
 \nabla_j  \delta ( \vec x-\vec X_0 )\;
{\mathsf  N}_{mlk} \nabla_l \right|_{\vec x=\vec X_0} \, ,
\label{potential}
\end{equation}
with
\beq
\label{calA}
  {\cal A}  \equiv  \frac{8}{\pi^2}\frac{(\mu b)^2}{m}  \frac{L}{ \left( \omega^2 - (\pi c/L)^2 - i\omega B/m\right)} \, , \hspace{1em} c^2 \equiv {\cal T}/m
  \eeq
 
 The next step is to think of the right-hand-side of (\ref{EQondesV}) to consist, not of a single string, but of many strings, randomly located and oriented, and with a distribution of lengths given by $p(L)$. In this case the problem is of a wave traveling in a random medium. Or, more precisely, in a medium filled with random scatterers. In any case, it is possible to find a coherent wave solution that is described by a complex index of refraction. This is achieved through the computation of an effective, average, Green's function. 
 
 Consider the full Green's function for the problem (\ref{EQondesV}):
 \bea
 \rho \omega^2 G_{im}(\vec x,\omega)+c_{ijkl} \nabla_j \nabla_l G_{km}  (\vec x,\omega)=  \hspace{1em} & & \nonumber \\
 \hspace{1em} - \sum_{\text{strings}} V_{ik}G_{km}(\vec x,\omega) -
  \delta_{im} \delta(\vec x)  \, . & &
\label{EqOndemodif}
\eea
To this end the average Green's function is written as (``Dyson's equation'')
\begin{equation}
 \langle \langle G\rangle \rangle = \left[ (G^0)^{-1}-\Sigma \right]^{-1}
\label{dyson}
\end{equation}
where $G^0$ is the Green's function of the medium without dislocations (i.e., ``free'', or ``bare'') {\cite{Maurel2005b,Weaver1990,Maurel2004}}, for an infinite, homogeneous and isotropic medium, and the double brackets $\langle \langle \cdot \rangle \rangle$ denote an average over the random variables characterizing the disordered medium, not to be confused with the thermal average considered in the text. The mass operator $\Sigma$ can be computed from the interaction $V$ according to a well-defined, perturbative procedure. It turns out that, due to the point-like nature of the interaction in (\ref{EqOndemodif}), the perturbation series can be summed and the result is
\beq
\label{massopstrings}
\Sigma_{ij} = -\int \sigma_0(L) p(L) dL \left[ \frac 15 (\delta_{ij} - \hat k_i \hat k_j ) k^2 + \frac{4}{15} k_ik_j\right]
\eeq
with
\beq
\label{sigma0}
\sigma_0 \equiv \frac{2\cal A}{1+ {\cal A}I}  \,  \hspace{1em} \mbox{and} \hspace{1em} I \equiv -i \om^3 \frac{1}{5 \pi} \frac{1}{\rho c_T^5} \, .
\eeq 

Clearly, just like $G^0$, the mass operator $\Sigma$, and consequently the effective Green's function $ \langle \langle G\rangle \rangle$, split into longitudinal and transverse parts:
\beq
\langle \langle G_{ij} \rangle \rangle =  \langle \langle G_L\rangle \rangle \hat k_i \hat k_j + \langle \langle G_T \rangle \rangle (\delta_{ij} - \hat k_i \hat k_j )
\eeq
 For the purposes of the present work we are only interested in the longitudinal portion and, from (\ref{dyson}), the result is
\beq
\langle \langle G_L \rangle \rangle^{-1} = \rho (- \om^2 +c_L^2 k^2 ) + k^2 \frac{4}{15} \int \sigma_0(\om,L) p(L) dL
\label{eq:Gsigma}
\eeq
Comparing (\ref{eq:fullgxi}) and (\ref{eq:Gsigma}) we obtain
\beq
f(\om) = - \frac{4}{15} \frac{1}{\rho c_L^2} \int \sigma_0(\om,L) p(L) dL
\label{eq:fandp(L)}
\eeq
which relates a population of strings with a distribution of lengths and the index of refraction of Section \ref{sec:addstrings}.

Putting together (\ref{calA}), (\ref{sigma0}) and (\ref{eq:fandp(L)}) we have
\beq
f(\om) = - f_0 \int \frac{L p(L) dL}{\om^2 -(\pi c/L)^2 -i\left( \om B/m + 2\om^3/(\alpha \om_{T}^2) \right)}
\label{eq:finalfom}
\eeq
where 
\beq
\label{eq:fzero}
f_0 \equiv  \frac 1\alpha \frac{8}{15} \frac{8}{\pi^2} \frac{c_T^4}{c_L^2} \, , \hspace{1em} \om_{T}^2 \equiv \frac{5\pi^3}{4} \frac{c_T}{L}
\eeq
and $\alpha$ is a dimensionless parameter of order one defined by $m \equiv \alpha \rho b^2$.

\section{The Damped Harmonic Oscillator Model (DHO)}
\label{sec:DHO}
{We follow the book of Boon-Yip \cite{BY}, see also Scopigno et al. \cite{Scopigno2005}. Starting from a linear description of a Navier-Stokes fluid with constant transport coefficient one finds, neglecting temperature fluctuations, that the current $j(\vec x,t)$ obeys the same Eqn. (\ref{eq:we_nostrings}) that has been our starting point for waves in an elastic solid, and, unsurprisingly, the current correlation is given by (\ref{eq:barecorrel1}). That is, Navier-Stokes sound in a fluid and elastic sound  in a continuous, homogeneous, solid obey the same equations at low frequency and wavenumber.
}

{In generalized hydrodynamics, the next step is to introduce a memory function into the equation for the current correlation function. The Navier-Stokes expression 
\beq
\frac{\partial J_{\rm NS}(k,t)}{\partial t} = -k^2 c_L^2 \int_0^t dt' J_{\rm NS}(k,t) -\nu k^2 J_{\rm NS}(k,t) \, ,
\label{JNS}
\eeq
where $\nu$ is the viscosity, is replaced by 
\bea
\frac{\partial J_{\rm GH}(k,t)}{\partial t}& = &-\frac{(k v_0)^2}{S(k)} \int_0^t dt' J_{\rm GH}(k,t)  \\
 && \hspace{2em} - k^2  \int_0^t dt' \Phi (k, t-t')J_{\rm GH}(k,t) \, . \nonumber
\label{JGH1}
\eea
Here, $v_0$ is the thermal particle velocity and $S(k)$ is the static structure factor. These quantities are determined by the material at hand, and the replacement of $c_L$ by $v_0/S(k)$ is dictated by one of the sum rules that must be obeyed by the current correlation function $J(k,t)$. The ``memory function'' $\Phi(k,t)$ is, as yet, undetermined, and is supposed to capture the fact that, at high frequencies and short wavelengths, viscous effects are no longer described by constant coefficients, but depend on time and length scales. The current correlation obtained from this expression is
\bea
\label{JGH1FT}
J_{\rm GH}(k,\om) &=& \\
&& \hspace{-4.5em}
 \frac{2v_0^2 \om^2 k^2 \Phi'(k,\om)}{[\om^2 -[(kv_0)^2/S(k)] +\om k^2 \Phi''(k,\om)]^2 +[\om k^2 \Phi' (k,\om) ]^2} \nonumber \, .
\eea
where $\Phi'$ and $\Phi''$ are the real and imaginary parts of the Laplace-Fourier transform of $\Phi(k,t)$.
}

{The ``Damped Harmonic Oscillator" (DHO) expression employed in Refs. \cite{Monaco2009} and \cite{Baldi2011} to interpret the glycerol and silica data is obtained from (\ref{JGH1FT}) with the special choices
\beq
\Phi'' (k,\om) = 0 \, , \hspace{.5em} k^2 \Phi' (k,\om) \equiv 2 \Gamma(k) \hspace{.5em}  \mbox{and} \hspace{.5em} \frac{(kv_0)^2}{S(k)} \equiv \Omega^2(k) \, .
\label{GH2DHO}
\eeq
That is, the damping is instantaneous (i.e., independent of frequency) but wavenumber-dependent. The functions $\Gamma(k)$ and $\Omega(k)$ are obtained through a comparison with the experimental data, and they do not appear to be directly related to a specific particle behavior in the glass. They do reproduce, however, the Navier-Stokes behavior at long wavelengths.
}

{In the body of this paper we have used an expression for the current correlation given by Eqs. (\ref{eq:mainJg}), (\ref{eq:fullgxi}) and (\ref{eq:finalfom}). This expression differs from a DHO model. It can however, be obtained as a special case of the generalized hydrodynamics expression, Eq. (\ref{JGH1FT}), with
\bea
\label{mem2string1}
\om \Phi' &=& c_L^2 (\xi \om + \text{Im} f) \\
 \frac{v_0^2}{S(k)} -\om \Phi''  & = & c_L^2( 1 -\text{Re}f ) \, .
\label{mem2string2}
\eea
These two formulae show that the string model developed in the body of the paper can be considered as being encompassed by a memory function formalism, albeit with a somewhat roundabout formulation. The coherent-wave-picture developed in Section \ref{Sec:string_model} provides a more direct formulation, that in addition has a straightforward physical interpretation, and it links the wave behavior to the vibrational degrees of freedom contained in the Boson peak, through the function $f(\om)$. These coherent waves are characterized by a phase velocity and attenuation given by Eqs. (\ref{phasevel}) and (\ref{attenuationGamma}), which differ from the position divided by wave vector and full width at half-maximum of the inelastic peak in the dynamic structure function, which are the definitions of velocity and attenuation commonly associated with the DHO model. Both the string and the DHO models provide adequate fits of the experimental data. They differ, however, in terms of physical interpretation.
}

\end{appendix}

%

\end{document}